# A novel two-party semiquantum key distribution protocol based on GHZ-like states

Tian-Jie Xu, Tian-Yu Ye*

College of Information & Electronic Engineering, Zhejiang Gongshang University, Hangzhou 310018, P.R.China

E-mail：yetianyu@mail.zjgsu.edu.cn

**Abstract:** In this paper, we propose a novel two-party semiquantum key distribution (SQKD) protocol by only employing one kind of GHZ-like state. The proposed SQKD protocol can create a private key shared between one quantum party with unlimited quantum abilities and one classical party with limited quantum abilities without the existence of a third party. The proposed SQKD protocol doesn't need the Hadamard gate or quantum entanglement swapping. Detailed security analysis turns out that the proposed SQKD protocol can resist various famous attacks from an outside eavesdropper, such as the Trojan horse attacks, the entangle-measure attack, the double CNOT attacks, the measure-resend attack and the intercept-resend attack.

## 1 Introduction

In 1984, Bennett and Brassard [1] proposed the first quantum key distribution (QKD) protocol and quickly attracted people's attention into the field of quantum cryptography. Classical cryptography can only ensure its security via the computational complexity of solving mathematical problems. Fortunately, quantum cryptography can gain its theoretically unconditional security based on the laws of quantum mechanics. Because of the advantage of unconditional security, various types of quantum cryptography protocols have been proposed, such as quantum key agreement (QKA) protocols [2-5], quantum dialogue (QD) protocols [6-19], *etc*.

In a QKD scheme, each party is required to possess full quantum abilities. Due to the high cost of quantum devices, it may be unpractical to ask all participants to have complete quantum capabilities in some realistic situations. In order to mitigate the burden of quantum state generation and measurement for a portion of participants, Boyer *et al.* [20,21] put forward two novel semiquantum key distribution (SQKD) protocols in 2007 and in 2009, respectively, which implied the birth of semiquantum cryptography. The above two protocols define that the classical party only has limited quantum abilities, such as preparing and measuring qubits in the classical basis $\{|0\rangle,|1\rangle\}$, sending qubits without disturbance and reordering qubits. In 2009, Zou *et al.* [22] proposed a number of SQKD protocols, each of which adopts less than four quantum states as the initial quantum resource. In 2011, Wang *et al.* [23] presented a two-party SQKD protocol by using one kind of Bell states. In 2013, Sun *et al.* [24] proposed two SQKD protocols without entanglement to exempt the classical party from measurement. In 2014, Yu *et al.* [25] put forward two authenticated SQKD protocols without using authenticated classical channels. In the same year, Krawec [26] designed a SQKD protocol which allows two classical users to create a shared key under the help of a server with full quantum capabilities. In 2018, Zhu *et al.* [27] proposed two SQKD protocols with three parties based on GHZ states. In 2020, Chen *et al.* [28] put forward two SQKD protocols with two parties and three parties, respectively, based on GHZ-like states. In 2020 and 2022, Ye *et al.* [29,30] proposed two novel SQKD protocols based on single photons in both polarization and spatial-mode degrees of freedom.

Based on the above analysis, in this paper, we propose a novel two-party SQKD protocol by only adopting one kind of GHZ-like state as the initial quantum resource, which realizes the establishment of a private shared key between one quantum user and one classical user without the existence of a third party. The proposed SQKD protocol is proven in detail to be secure.

The remaining parts of this paper are arranged as follows: Section 2 describes the proposed two-party SQKD protocol based on GHZ-like states; Section 3 analyzes its security; and finally, discussions and conclusions are delivered in Section 4.

## 2 The proposed two-party SQKD protocol based on GHZ-like states

The GHZ-type states [31] can be expressed as

$$|G_{000}\rangle = \frac{1}{2}(|000\rangle + |011\rangle + |101\rangle + |110\rangle),\tag{1}$$

$$|G_{001}\rangle = \frac{1}{2}(|001\rangle + |010\rangle + |100\rangle + |111\rangle),\tag{2}$$

$$|G_{010}\rangle = \frac{1}{2}(|000\rangle - |011\rangle - |101\rangle + |110\rangle),\tag{3}$$

$$|G_{011}\rangle = \frac{1}{2}(|001\rangle - |010\rangle - |100\rangle + |111\rangle),\tag{4}$$

$$|G_{100}\rangle = \frac{1}{2}(|000\rangle - |011\rangle + |101\rangle - |110\rangle),\tag{5}$$

$$|G_{101}\rangle = \frac{1}{2}(|001\rangle - |010\rangle + |100\rangle - |111\rangle),\tag{6}$$

$$|G_{110}\rangle = \frac{1}{2}(|000\rangle + |011\rangle - |101\rangle - |110\rangle),\tag{7}$$

$$|G_{111}\rangle = \frac{1}{2}(|001\rangle + |010\rangle - |100\rangle - |111\rangle).\tag{8}$$

The set of $\{|G_{abc}\rangle\,|\,a,b,c\in\{0,1\}\}$, which can be named as the GHZ-like basis, makes up an orthonormal basis for the space of a tripartite quantum system.

The quantum resource used in the proposed two-party SQKD protocol is the GHZ-like state $|G_{001}\rangle_{123}$, which can also be expressed as

$$|G_{001}\rangle_{123} = \frac{1}{\sqrt{2}}(|0\rangle_1|\psi^+\rangle_{23} + |1\rangle_1|\phi^+\rangle_{23}),\tag{9}$$

where $|\psi^+\rangle = \frac{1}{\sqrt{2}}(|01\rangle + |10\rangle)$ and $|\phi^+\rangle = \frac{1}{\sqrt{2}}(|00\rangle + |11\rangle)$. Further, we define two unitary operations $\sigma_0$ and $\sigma_1$ as follows:

$$\sigma_0 = |0\rangle\langle0| + |1\rangle\langle1|,\tag{10}$$

$$\sigma_1 = |0\rangle\langle1| + |1\rangle\langle0|.\tag{11}$$

Suppose that Alice has unlimited quantum abilities, while Bob is the classical party with limited quantum capabilities. Alice and Bob want to share a secure key via the following procedures.

**Step 1:** Alice prepares $4(n+\delta+\nu)$ GHZ-like states all in the state of $|G_{001}\rangle$, where $n$, $\delta$ and $\nu$

are positive integers. Then, Alice divides these GHZ-like states into three particle sequences: $S_l = \left\{ S_l^1, S_l^2, \cdots, S_l^t, \cdots, S_l^{4(n+\delta+\nu)} \right\}$, where $S_l$ is the sequence composed by all $l$ th particles of these $4(n+\delta+\nu)$ GHZ-like states, $S_l^t$ is the $t$ th particle of $S_l$, $l = 1,2,3$ and $t = 1,2,\cdots,4(n+\delta+\nu)$. Alice keeps the particles of $S_2$ and $S_3$ in her hand, and sends the particles of $S_1$ to Bob through the quantum channel one by one. Note except the first particle of $S_1$, Alice sends out the next one only after the previous one has arrived at her site.

**Step 2:** Bob randomly executes the CTRL operations or the SIFT operations on the particles in $S_1$. The CTRL operation means to reflect the $t$ th received particle to Alice without disturbance, while the SIFT operation means to measure the $t$ th received particle with the $Z$ basis (i.e., the $\{|0\rangle, |1\rangle\}$ basis), prepare a fresh particle in the state as found and resend it to Alice. Here, $t = 1,2,\cdots,4(n+\delta+\nu)$.

**Step 3:** After Alice receives all particles in $S_1$ after Bob's operations, Bob announces the positions of the particles in $S_1$ where he chose to perform the CTRL operations.

For the sake of security check on the CTRL particles, for the position where Bob executed the CTRL operation, Alice measures the particle of $S_1$ from Bob and the corresponding particles in $S_2$ and $S_3$ at her site with the GHZ-like basis. If Eve is not online, Alice's measurement result should be $|G_{001}\rangle$.

For the sake of security check on the SIFT particles, Alice randomly selects $2\delta$ positions where Bob executed the SIFT operations and requires Bob to tell her his corresponding measurement results on these chosen positions. For each of these chosen positions, Alice measures the particle of $S_1$ from Bob with the $Z$ basis and measures the corresponding particles in $S_2$ and $S_3$ at her site with the Bell basis. If Eve is not online, Bob's measurement result on the particle of $S_1$ on the chosen position, Alice's measurement result on the particle of $S_1$ on the chosen position and Alice's measurement result on the corresponding particles in $S_2$ and $S_3$ at her site on the chosen position should be correctly correlated.

If no Eve is found in the end, the protocol will be continued to the next step; otherwise, it will be terminated.

**Step 4:** Alice discards all particles in $S_1$, $S_2$ and $S_3$ on the $2(n+\delta+\nu)$ positions where Bob performed the CTRL operations and all particles in $S_1$, $S_2$ and $S_3$ on the $2\delta$ positions chosen for the security check on SIFT particles. For simplicity, the two ordered sequences composed by the remaining particles in $S_2$ and $S_3$ at Alice's site are called as $S_2'$ and $S_3'$, respectively, while the ordered sequence formed by the particles in $S_1$ corresponding to the ones in $S_2'$ and $S_3'$ is named as $S_1'$. Since the particles in $S_1'$ were prepared by Bob, he can automatically know their states. The state of the $i$ th particle in $S_1'$ is coded as $M_{B1}^i$, where $i = 1,2,\cdots,2(n+\nu)$. Concretely speaking, if the state of the $i$ th particle in $S_1'$ is $|0\rangle$, then $M_{B1}^i = 0$; and if the state of the $i$ th particle in $S_1'$ is $|1\rangle$, then $M_{B1}^i = 1$. Then, Alice measures the particles in $S_1'$ with the $Z$ basis. Alice can deduce out the state of the $i$ th particles in $S_2'$ and $S_3'$ from her measurement result on the $i$ th corresponding particle in $S_1'$. Concretely speaking, if her measurement result on the $i$ th particle in $S_1'$ is $|0\rangle$, the state of the $i$ th particles in $S_2'$ and $S_3'$ will be $|\psi^+\rangle$ which can be coded as $M_{A1}^i = 0$; and if her measurement result on the $i$ th particle in $S_1'$ is $|1\rangle$), the state of the $i$ th particles in $S_2'$ and $S_3'$ will be $|\phi^+\rangle$ which can be coded as $M_{A1}^i = 1$. Let $M_{A1} = \left\{ M_{A1}^1, M_{A1}^2, \cdots, M_{A1}^{2(n+\nu)} \right\}$ and $M_{B1} = \left\{ M_{B1}^1, M_{B1}^2, \cdots, M_{B1}^{2(n+\nu)} \right\}$. It is apparent that $M_{A1} = M_{B1}$.

**Step 5:** According to the states of the particles in $S_2'$ and $S_3'$, Alice performs the corresponding unitary operations on the particles of $S_2'$ to obtain the new sequence $S_2'^U$. The rule is that: when the state of the $i$ th particles in $S_2'$ and $S_3'$ is $|\phi^+\rangle$, Alice performs $\sigma_0$ on the $i$ th particle in $S_2'$; and when the state of the $i$ th particles in $S_2'$ and $S_3'$ is $|\psi^+\rangle$, Alice performs $\sigma_1$ on the $i$ th particle in $S_2'$. Here, $i = 1, 2, \cdots, 2(n+\nu)$. Subsequently, Alice sends the particles of $S_2'^U$ to Bob via quantum channel one by one. Note except the first particle of $S_2'^U$, Alice sends out the next one only after the previous one has arrived at her site.

**Step 6:** Bob randomly executes the CTRL operations or the SIFT operations on the particles in $S_2'^U$.

**Step 7:** After Alice receives all particles in $S_2'^U$ after Bob's operations, Bob announces on which particles in $S_2'^U$ he chose to perform the CTRL operations.

For the sake of security check on the CTRL particles, for the position where Bob executed the CTRL operation, Alice measures the particle of $S_2'^U$ from Bob and the corresponding particle in $S_3'$ at her site with the Bell basis. If Eve is not online, Alice's measurement result should be $|\phi^+\rangle$.

For the sake of security check on the SIFT particles, Alice randomly selects $\nu$ positions where Bob executed the SIFT operations and requires Bob to tell her his corresponding measurement results on these chosen positions. For each of these chosen positions, Alice measures the particle of $S_2'^U$ from Bob and the corresponding particle in $S_3'$ at her site with the $Z$ basis. If Eve is not online, Bob's measurement result on the particle of $S_2'^U$ on the chosen position, Alice's measurement result on the particle of $S_2'^U$ on the chosen position and Alice's measurement result on the corresponding particle in $S_3'$ at her site on the chosen position should be same.

If no Eve is found in the end, the protocol will be continued to the next step; otherwise, it will be terminated.

**Step 8:** Alice discards all particles in $S_2'^U$ and $S_3'$ on the $n+\nu$ positions where Bob performed the CTRL operations and all particles in $S_2'^U$ and $S_3'$ on the $\nu$ positions chosen for the security check on SIFT particles. For simplicity, the two ordered sequences composed by the remaining particles in $S_2'^U$ and $S_3'$ at Alice's site are called as $S_2''^U$ and $S_3''$, respectively. Since the particles in $S_2''^U$ were prepared by Bob, he can automatically know their states. The state of the $k$ th particle in $S_2''^U$ is coded as $M_{B2}^k$, where $k = 1, 2, \cdots, n$. Concretely speaking, if the state of the $k$ th particle in $S_2''^U$ is $|0\rangle$, then $M_{B2}^k = 0$; and if the state of the $k$ th particle in $S_2''^U$ is $|1\rangle$, then $M_{B2}^k = 1$. Then, Alice measures the particles in $S_3''$ with the $Z$ basis. If her measurement result on the $k$ th particle in $S_3''$ is $|0\rangle$, then $M_{A2}^k = 0$; and if her measurement result on the $k$ th particle in $S_3''$ is $|1\rangle$, then $M_{A2}^k = 1$. Let $M_{A2} = \left\{ M_{A2}^1, M_{A2}^2, \cdots, M_{A2}^n \right\}$ and $M_{B2} = \left\{ M_{B2}^1, M_{B2}^2, \cdots, M_{B2}^n \right\}$. It is apparent that $M_{A2} = M_{B2}$.

**Step 9:** Alice and Bob accept $M_{A1} \| M_{A2}$ and $M_{B1} \| M_{B2}$ as the INFO bits, respectively, where $\|$ is the symbol of concatenation. It is apparent that $M_{A1} \| M_{A2} = M_{B1} \| M_{B2}$. Alice and Bob can derive the $m$-bit final key from the $3n + 2v$ INFO bits by using error correction code (ECC) and privacy amplification (PA).

## 3 Security analysis

In this section, we conduct security analysis on the transmissions of $S_1$ and $S_2^{,U}$, respectively.

### 3.1 Security analysis on the transmission of $S_1$

(1) The Trojan horse attacks

In this protocol, $S_1$ undergoes a circular transmission between Alice and Bob, so actions should be taken to prevent the invisible photon eavesdropping attack [32] and the delay-photon Trojan horse attack [33,34] from Eve. According to Refs.[34,35], a wavelength filter can be utilized by Bob to avoid the invisible photon eavesdropping attack; and a photon number splitter (PNS: 50/50) can be employed by Bob to resist the delay-photon Trojan horse attack.

(2) The entangle-measure attack

The entangle-measure attack launched by Eve on the particles of $S_1$ is shown in Fig.1. Eve imposes two unitaries, $U_E$ and $U_F$, which share a common probe space with the initial state $|\zeta\rangle_E$, on the particles of $S_1$, where $U_E$ attacks the particles from Alice to Bob and $U_F$ attacks the particles back to Alice. By virtue of the shared probe, Eve can launch her attack on the particles back to Alice depending on the information gained from $U_E$ ( if Eve does not exploit this fact, the "shared probe" can be thought to be the composite system of two independent probes) [20,21].

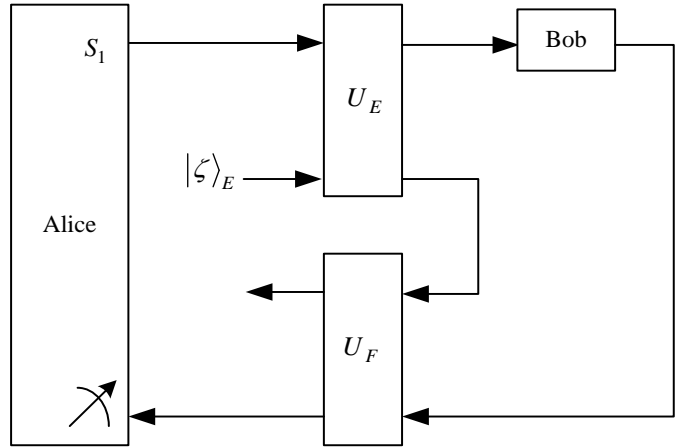

Fig.1 Eve's entangle-measure attack on the particles of $S_1$ with two unitaries $U_E$ and $U_F$

**Theorem 1.** *Suppose that Eve launches attack* $(U_E, U_F)$ *on the particle of* $S_1$ *from Alice to Bob and back to Alice. In order to introduce no error in Step 3, the final state of Eve's probe should be independent from both Bob's operation and Alice and Bob's measurement results.*

**Proof.** The effect of $U_E$ on the qubits $|0\rangle$ and $|1\rangle$ can be described as

$$U_E(|0\rangle|\zeta\rangle_E) = \beta_{00}|0\rangle|\zeta_{00}\rangle + \beta_{01}|1\rangle|\zeta_{01}\rangle, \tag{12}$$

$$U_E(|1\rangle|\zeta\rangle_E) = \beta_{10}|0\rangle|\zeta_{10}\rangle + \beta_{11}|1\rangle|\zeta_{11}\rangle, \tag{13}$$

where $|\zeta_{00}\rangle$, $|\zeta_{01}\rangle$, $|\zeta_{10}\rangle$ and $|\zeta_{11}\rangle$ are Eve's probe states determined by $U_E$, $|\beta_{00}|^2 + |\beta_{01}|^2 = 1$ and

$\left|\beta_{10}\right|^2 + \left|\beta_{11}\right|^2 = 1$ .

According to Stinespring dilation theorem, the global state of the composite system before Bob's operation is

$$U_E\left(\left|G_{001}\right\rangle_{123}\left|\zeta\right\rangle_E\right) = U_E\left[\frac{1}{2}\left(\left|001\right\rangle + \left|010\right\rangle + \left|100\right\rangle + \left|111\right\rangle\right)_{123}\left|\zeta\right\rangle_E\right]$$

$$= \frac{1}{2}\Big[\left(\beta_{00}\left|0\right\rangle_1\left|\zeta_{00}\right\rangle + \beta_{01}\left|1\right\rangle_1\left|\zeta_{01}\right\rangle\right)\left|0\right\rangle_2\left|1\right\rangle_3 + \left(\beta_{00}\left|0\right\rangle_1\left|\zeta_{00}\right\rangle + \beta_{01}\left|1\right\rangle_1\left|\zeta_{01}\right\rangle\right)\left|1\right\rangle_2\left|0\right\rangle_3$$

$$+ \left(\beta_{10}\left|0\right\rangle_1\left|\zeta_{10}\right\rangle + \beta_{11}\left|1\right\rangle_1\left|\zeta_{11}\right\rangle\right)\left|0\right\rangle_2\left|0\right\rangle_3 + \left(\beta_{10}\left|0\right\rangle_1\left|\zeta_{10}\right\rangle + \beta_{11}\left|1\right\rangle_1\left|\zeta_{11}\right\rangle\right)\left|1\right\rangle_2\left|1\right\rangle_3\Big]$$

$$= \frac{1}{2}\Big(\left|0\right\rangle_1\left|0\right\rangle_2\left|1\right\rangle_3\beta_{00}\left|\zeta_{00}\right\rangle + \left|1\right\rangle_1\left|0\right\rangle_2\left|1\right\rangle_3\beta_{01}\left|\zeta_{01}\right\rangle + \left|0\right\rangle_1\left|1\right\rangle_2\left|0\right\rangle_3\beta_{00}\left|\zeta_{00}\right\rangle + \left|1\right\rangle_1\left|1\right\rangle_2\left|0\right\rangle_3\beta_{01}\left|\zeta_{01}\right\rangle$$

$$+ \left|0\right\rangle_1\left|0\right\rangle_2\left|0\right\rangle_3\beta_{10}\left|\zeta_{10}\right\rangle + \left|1\right\rangle_1\left|0\right\rangle_2\left|0\right\rangle_3\beta_{11}\left|\zeta_{11}\right\rangle + \left|0\right\rangle_1\left|1\right\rangle_2\left|1\right\rangle_3\beta_{10}\left|\zeta_{10}\right\rangle + \left|1\right\rangle_1\left|1\right\rangle_2\left|1\right\rangle_3\beta_{11}\left|\zeta_{11}\right\rangle\Big)$$

$$= \left|0\right\rangle_1\left|0\right\rangle_2\left|1\right\rangle_3\left|E_{00}\right\rangle + \left|1\right\rangle_1\left|0\right\rangle_2\left|1\right\rangle_3\left|E_{01}\right\rangle + \left|0\right\rangle_1\left|1\right\rangle_2\left|0\right\rangle_3\left|E_{00}\right\rangle + \left|1\right\rangle_1\left|1\right\rangle_2\left|0\right\rangle_3\left|E_{01}\right\rangle$$

$$+ \left|0\right\rangle_1\left|0\right\rangle_2\left|0\right\rangle_3\left|E_{10}\right\rangle + \left|1\right\rangle_1\left|0\right\rangle_2\left|0\right\rangle_3\left|E_{11}\right\rangle + \left|0\right\rangle_1\left|1\right\rangle_2\left|1\right\rangle_3\left|E_{10}\right\rangle + \left|1\right\rangle_1\left|1\right\rangle_2\left|1\right\rangle_3\left|E_{11}\right\rangle , \quad (14)$$

where the subscripts 1, 2 and 3 denote the particles from $S_1$, $S_2$ and $S_3$, respectively, and $\left|E_{00}\right\rangle = \frac{1}{2}\beta_{00}\left|\zeta_{00}\right\rangle$ , $\left|E_{01}\right\rangle = \frac{1}{2}\beta_{01}\left|\zeta_{01}\right\rangle$ , $\left|E_{10}\right\rangle = \frac{1}{2}\beta_{10}\left|\zeta_{10}\right\rangle$ and $\left|E_{11}\right\rangle = \frac{1}{2}\beta_{11}\left|\zeta_{11}\right\rangle$ .

When Bob receives the particle of $S_1$ from Alice, he performs either the SIFT operation or the CTRL operation on it. Subsequently, Eve performs $U_F$ on the particle back to Alice.

(i) Case 1: Bob chooses the SIFT operation. As a result, the state of the composite system of Eq.(14) is collapsed into $\left|0\right\rangle_1\left|0\right\rangle_2\left|1\right\rangle_3\left|E_{00}\right\rangle + \left|0\right\rangle_1\left|1\right\rangle_2\left|0\right\rangle_3\left|E_{00}\right\rangle + \left|0\right\rangle_1\left|0\right\rangle_2\left|0\right\rangle_3\left|E_{10}\right\rangle + \left|0\right\rangle_1\left|1\right\rangle_2\left|1\right\rangle_3\left|E_{10}\right\rangle$ , when Bob's measurement result is $\left|0\right\rangle_1$ ; or $\left|1\right\rangle_1\left|0\right\rangle_2\left|1\right\rangle_3\left|E_{01}\right\rangle + \left|1\right\rangle_1\left|1\right\rangle_2\left|0\right\rangle_3\left|E_{01}\right\rangle + \left|1\right\rangle_1\left|0\right\rangle_2\left|0\right\rangle_3\left|E_{11}\right\rangle + \left|1\right\rangle_1\left|1\right\rangle_2\left|1\right\rangle_3\left|E_{11}\right\rangle$ , when Bob's measurement result is $\left|1\right\rangle_1$ .

Eve performs $U_F$ on the particle back to Alice. In order that Eve induces no error on the SIFT particle in Step 3, the state of the composite system should be turned into

$$U_F\left(\left|0\right\rangle_1\left|0\right\rangle_2\left|1\right\rangle_3\left|E_{00}\right\rangle + \left|0\right\rangle_1\left|1\right\rangle_2\left|0\right\rangle_3\left|E_{00}\right\rangle + \left|0\right\rangle_1\left|0\right\rangle_2\left|0\right\rangle_3\left|E_{10}\right\rangle + \left|0\right\rangle_1\left|1\right\rangle_2\left|1\right\rangle_3\left|E_{10}\right\rangle\right) = \left|0\right\rangle_1\left|\psi^+\right\rangle_{23}\left|F_0\right\rangle , \quad (15)$$

when Bob's measurement result is $\left|0\right\rangle_1$ ; or

$$U_F\left(\left|1\right\rangle_1\left|0\right\rangle_2\left|1\right\rangle_3\left|E_{01}\right\rangle + \left|1\right\rangle_1\left|1\right\rangle_2\left|0\right\rangle_3\left|E_{01}\right\rangle + \left|1\right\rangle_1\left|0\right\rangle_2\left|0\right\rangle_3\left|E_{11}\right\rangle + \left|1\right\rangle_1\left|1\right\rangle_2\left|1\right\rangle_3\left|E_{11}\right\rangle\right) = \left|1\right\rangle_1\left|\phi^+\right\rangle_{23}\left|F_1\right\rangle , \quad (16)$$

when Bob's measurement result is $\left|1\right\rangle_1$ .

(ii) Case 2: Bob chooses the CTRL operation. As a result, after Eve performs $U_F$ on the particle back to Alice, due to Eq.(15) and Eq.(16), the state of the composite system is evolved into

$$U_F\left[U_E\left(\left|G_{001}\right\rangle_{123}\left|\zeta\right\rangle_E\right)\right] = U_F\left(\left|0\right\rangle_1\left|0\right\rangle_2\left|1\right\rangle_3\left|E_{00}\right\rangle + \left|1\right\rangle_1\left|0\right\rangle_2\left|1\right\rangle_3\left|E_{01}\right\rangle + \left|0\right\rangle_1\left|1\right\rangle_2\left|0\right\rangle_3\left|E_{00}\right\rangle + \left|1\right\rangle_1\left|1\right\rangle_2\left|0\right\rangle_3\left|E_{01}\right\rangle$$

$$+ \left|0\right\rangle_1\left|0\right\rangle_2\left|0\right\rangle_3\left|E_{10}\right\rangle + \left|1\right\rangle_1\left|0\right\rangle_2\left|0\right\rangle_3\left|E_{11}\right\rangle + \left|0\right\rangle_1\left|1\right\rangle_2\left|1\right\rangle_3\left|E_{10}\right\rangle + \left|1\right\rangle_1\left|1\right\rangle_2\left|1\right\rangle_3\left|E_{11}\right\rangle\right)$$

$$= \left|0\right\rangle_1\left|\psi^+\right\rangle_{23}\left|F_0\right\rangle + \left|1\right\rangle_1\left|\phi^+\right\rangle_{23}\left|F_1\right\rangle . \quad (17)$$

In order that Eve induces no error on the CTRL particle in Step 3, Alice's GHZ-like basis measurement result on the particle in $S_1$ reflected by Bob and the corresponding particles in $S_2$ and

$S_3$ at her site should be in the state of $|G_{001}\rangle_{123}$. Thus, we should have

$$|F_0\rangle = |F_1\rangle = |F\rangle. \tag{18}$$

(iii) Inserting Eq.(18) into Eq.(15) derives

$$U_F\left(|0\rangle_1|0\rangle_2|1\rangle_3|E_{00}\rangle + |0\rangle_1|1\rangle_2|0\rangle_3|E_{00}\rangle + |0\rangle_1|0\rangle_2|0\rangle_3|E_{10}\rangle + |0\rangle_1|1\rangle_2|1\rangle_3|E_{10}\rangle\right) = |0\rangle_1|\psi^+\rangle_{23}|F\rangle. \tag{19}$$

Inserting Eq.(18) into Eq.(16) produces

$$U_F\left(|1\rangle_1|0\rangle_2|1\rangle_3|E_{01}\rangle + |1\rangle_1|1\rangle_2|0\rangle_3|E_{01}\rangle + |1\rangle_1|0\rangle_2|0\rangle_3|E_{11}\rangle + |1\rangle_1|1\rangle_2|1\rangle_3|E_{11}\rangle\right) = |1\rangle_1|\phi^+\rangle_{23}|F\rangle. \tag{20}$$

Inserting Eq.(18) into Eq.(17) produces

$$U_F\left[U_E\left(|G_{001}\rangle_{123}|\zeta\rangle_E\right)\right] = |G_{001}\rangle_{123}|F\rangle. \tag{21}$$

It can be concluded from Eqs.(19-21) that in order to introduce no error in Step 3, the final state of Eve's probe should be independent from both Bob's operation and Alice and Bob's measurement results.

(3) The double CNOT attacks

In Step 1, Alice keeps the particles of $S_2$ and $S_3$ in her hand, and sends Bob the particles of $S_1$ one by one. Eve intercepts the particle from Alice to Bob and executes the first CNOT attack on her own auxiliary qubit $|0\rangle_E$ and it, where the former and the latter are the target qubit and the control qubit, respectively. After Eve performs the first CNOT operation, the global state of the whole system can be expressed as

$$CNOT_{1E}\left(|G_{001}\rangle_{123}\otimes|0\rangle_E\right) = \frac{1}{2}\left(|0010\rangle + |0100\rangle + |1001\rangle + |1111\rangle\right)_{123E}. \tag{22}$$

After the first CNOT operation, Eve sends the particle of $S_1$ to Bob. In Step 2, Bob randomly performs the CTRL operation or the SIFT operation on the received particle. Eve intercepts the particle returned by Bob. As Eve cannot discriminate Bob's operation, in order to escape from being detected, Eve has to carry out the second CNOT attack, where the particle from Bob and her auxiliary qubit are the control qubit and the target qubit, respectively. There are two cases to be discussed.

Case 1: Bob chooses the CTRL operation. After Eve performs the second CNOT attack, the global state of the whole system can be expressed as

$$CNOT_{1E}^{\otimes 2}\left(|G_{001}\rangle_{123}\otimes|0\rangle_E\right) = \frac{1}{2}\left(|0010\rangle + |0100\rangle + |1000\rangle + |1110\rangle\right)_{123E} = |G_{001}\rangle_{123}|0\rangle_E. \tag{23}$$

Case 2: Bob chooses the SIFT operation. As a result, the global state of the whole system in Eq.(22) will be collapsed into $\left(|0010\rangle + |0100\rangle\right)_{123E}$ or $\left(|1001\rangle + |1111\rangle\right)_{123E}$. After Eve performs the second CNOT attack, the global state of the whole system is changed into

$$CNOT_{1E}\left(|0010\rangle + |0100\rangle\right)_{123E} = \sqrt{2}|0\rangle_1|\psi^+\rangle_{23}|0\rangle_E, \quad \text{if Bob's measurement result is } |0\rangle_1; \tag{24}$$

$$CNOT_{1E}\left(|1001\rangle + |1111\rangle\right)_{123E} = \sqrt{2}|1\rangle_1|\phi^+\rangle_{23}|0\rangle_E, \quad \text{if Bob's measurement result is } |1\rangle_1. \tag{25}$$

From Eqs.(23-25), it is easy to know that although Eve can successfully pass the first eavesdropping detection in Step 3 after her double CNOT attacks, he still can't know both Bob's operation and Alice and Bob's measurement results, because the state of Eve's auxiliary particle after her double CNOT attacks is always $|0\rangle_E$. Thus, this protocol can resist this kind of double CNOT attacks.

(4) The measure-resend attack

In order to try to obtain $M_{B1}$, Eve intercepts the particles in $S_1$ from Alice to Bob and uses the $Z$ basis to measure them. In this way, each of the particles in $S_1$ is randomly collapsed into $|0\rangle_1$ or $|1\rangle_1$ with the same probability. Without losing generality, suppose that the $t$ th particle in $S_1$ after Eve's measurement is collapsed into $|0\rangle_1$. When Bob chooses the CTRL operation, after Alice hears of Bob's operation, Alice uses the GHZ-like basis to measure the $t$ th particle of $S_1$ from Bob and the corresponding $t$ th particles in $S_2$ and $S_3$ in her hand; in this way, the probability that Eve is detected by Alice and Bob is $\frac{1}{2}$ in this case, since Alice's GHZ-like basis measurement result is randomly in $|G_{001}\rangle_{123}$ or $|G_{111}\rangle_{123}$. When Bob chooses the SIFT operation, after Alice hears of Bob's operation, Alice will measure the $t$ th particle of $S_1$ from Bob with the $Z$ basis and the corresponding $t$ th particles in $S_2$ and $S_3$ at her site with the Bell basis if this position is chosen for the first eavesdropping detection; in this way, the probability that Eve is detected by Alice and Bob is 0 in this case, since Alice's measurement results on the $t$ th particle of $S_1$ from Bob and the corresponding $t$ th particles in $S_2$ and $S_3$ at her site are $|0\rangle_1$ and $|\psi^+\rangle_{23}$, respectively.

In summary, when Eve launches this measure-resend attack, the probability that Eve is detected by Alice and Bob in the first eavesdropping detection is $\frac{1}{2}\times\frac{1}{2}+\frac{1}{2}\times0=\frac{1}{4}$.

(5) The intercept-resend attack

In order to try to obtain $M_{B1}$, Eve produces the fake particles in the $Z$ basis, intercepts the particles in $S_1$ from Alice to Bob, and transmits the fake ones instead of that in $S_1$ to Bob. Without losing generality, assume that Eve prepares the $t$ th fake particle in the state of $|0\rangle_E$. In this way, Eve transmits the $t$ th fake particle $|0\rangle_E$ instead of the $t$ th particle in $S_1$ to Bob. When Bob chooses the CTRL operation, after Alice hears of Bob's operation, Alice uses the GHZ-like basis to measure the $t$ th fake particle $|0\rangle_E$ from Bob and the corresponding $t$ th particles in $S_2$ and $S_3$ in her hand; in this way, the probability that Eve is detected by Alice and Bob is $\frac{3}{4}$ in this case, as Alice's GHZ-like basis measurement result is randomly in $|G_{001}\rangle_{E23}$, $|G_{111}\rangle_{E23}$, $|G_{000}\rangle_{E23}$ or $|G_{110}\rangle_{E23}$. When Bob chooses the SIFT operation, after Alice hears of Bob's operation, Alice will measure the $t$ th fake particle $|0\rangle_E$ from Bob with the $Z$ basis and the corresponding $t$ th particles in $S_2$ and $S_3$ at her site with the Bell basis if this position is chosen for the first eavesdropping detection; as a result, the probability that Eve is detected by Alice and Bob is $\frac{1}{2}$ in this case, as Alice's measurement result on the corresponding $t$ th particles in $S_2$ and $S_3$ in her hand is randomly $|\psi^+\rangle_{23}$ or $|\phi^+\rangle_{23}$.

To sum up, when Eve launches the intercept-resend attack, the probability that Eve is detected by Alice and Bob in the first eavesdropping detection is $\frac{1}{2}\times\frac{3}{4}+\frac{1}{2}\times\frac{2\delta}{2(n+\delta+\nu)}\times\frac{1}{2}=\frac{3}{8}+\frac{\delta}{4(n+\delta+\nu)}$.

### 3.2 Security analysis on the transmission of $S_2'^U$

(1) The Trojan horse attacks

In this protocol, $S_2'^U$ undergoes a circular transmission between Alice and Bob. According to Refs.[34,35], Bob can utilize a wavelength filter and a photon number splitter (PNS: 50/50) to overcome the invisible photon eavesdropping attack and the delay-photon Trojan horse attack, respectively.

(2) The entangle-measure attack

The entangle-measure attack launched by Eve on the particles of $S_2'^U$ is shown in Fig.2. Eve imposes two unitaries, $U_H$ and $U_L$, which share a common probe space with the initial state $|\varepsilon\rangle_E$, on the particles of $S_2'^U$, where $U_H$ attacks the particles from Alice to Bob and $U_L$ attacks the particles back to Alice. By virtue of the shared probe, Eve can launch her attack on the particles back to Alice depending on the information gained from $U_H$ ( if Eve does not exploit this fact, the "shared probe" can be thought as the composite system of two independent probes) [20,21].

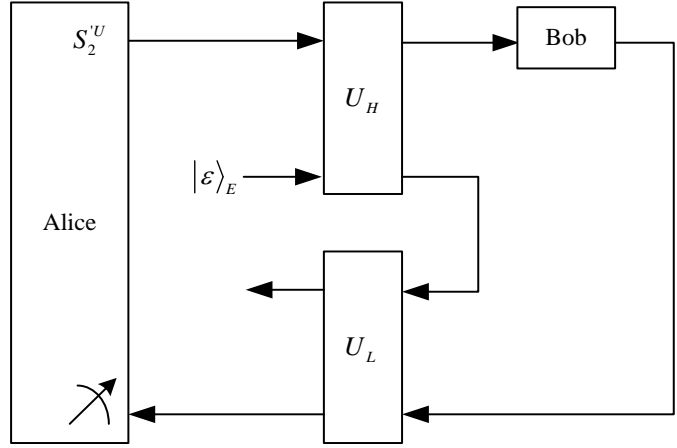

Fig.1 Eve's entangle-measure attack on the particles of $S_2'^U$ with two unitaries $U_H$ and $U_L$

**Theorem 1.** *Suppose that Eve launches attack $(U_H, U_L)$ on the particle of $S_2'^U$ from Alice to Bob and back to Alice. In order to introduce no error in Step 7, the final state of Eve's probe should be independent from both Bob's operation and Alice and Bob's measurement results.*

**Proof.** The effect of $U_H$ on the qubits $|0\rangle$ and $|1\rangle$ can be expressed as

$$U_H(|0\rangle|\varepsilon\rangle_E) = \alpha_{00}|0\rangle|\varepsilon_{00}\rangle + \alpha_{01}|1\rangle|\varepsilon_{01}\rangle, \tag{26}$$

$$U_H(|1\rangle|\varepsilon\rangle_E) = \alpha_{10}|0\rangle|\varepsilon_{10}\rangle + \alpha_{11}|1\rangle|\varepsilon_{11}\rangle, \tag{27}$$

where $|\varepsilon_{00}\rangle$, $|\varepsilon_{01}\rangle$, $|\varepsilon_{10}\rangle$ and $|\varepsilon_{11}\rangle$ are Eve's probe states determined by $U_H$, $|\alpha_{00}|^2 + |\alpha_{01}|^2 = 1$ and $|\alpha_{10}|^2 + |\alpha_{11}|^2 = 1$.

According to Stinespring dilation theorem, the global state of the composite system before Bob's operation is

$$U_H(|\phi^+\rangle_{23}|\varepsilon\rangle_E) = U_H\left[\frac{1}{\sqrt{2}}(|00\rangle + |11\rangle)_{23}|\varepsilon\rangle_E\right]$$

$$= \frac{1}{\sqrt{2}}\Big[\big(\alpha_{00}|0\rangle_2|\varepsilon_{00}\rangle + \alpha_{01}|1\rangle_2|\varepsilon_{01}\rangle\big)|0\rangle_3 + \big(\alpha_{10}|0\rangle_2|\varepsilon_{10}\rangle + \alpha_{11}|1\rangle_2|\varepsilon_{11}\rangle\big)|1\rangle_3\Big]$$

$$= \frac{1}{\sqrt{2}}\Big[|0\rangle_2\big(\alpha_{00}|0\rangle_3|\varepsilon_{00}\rangle + \alpha_{10}|1\rangle_3|\varepsilon_{10}\rangle\big) + |1\rangle_2\big(\alpha_{01}|0\rangle_3|\varepsilon_{01}\rangle + \alpha_{11}|1\rangle_3|\varepsilon_{11}\rangle\big)\Big], \tag{28}$$

where the subscripts 2 and 3 denote the particles from $S_2'^U$ and $S_3'$, respectively.

(i) Case 1: Bob chooses the SIFT operation. As a result, the global state of the composite

system in Eq.(28) is collapsed into $|0\rangle_2\left(\alpha_{00}|0\rangle_3|\varepsilon_{00}\rangle+\alpha_{10}|1\rangle_3|\varepsilon_{10}\rangle\right)$ when Bob's measurement result is $|0\rangle_2$; or $|1\rangle_2\left(\alpha_{01}|0\rangle_3|\varepsilon_{01}\rangle+\alpha_{11}|1\rangle_3|\varepsilon_{11}\rangle\right)$ when Bob's measurement result is $|1\rangle_2$.

Eve performs $U_L$ on the particle back to Alice. In order that Eve induces no error on the SIFT particle in Step 7, the state of the composite system should be turned into

$$U_L\left[|0\rangle_2\left(\alpha_{00}|0\rangle_3|\varepsilon_{00}\rangle+\alpha_{10}|1\rangle_3|\varepsilon_{10}\rangle\right)\right]=|0\rangle_2|0\rangle_3|\varepsilon_0\rangle,\tag{29}$$

when Bob's measurement result is $|0\rangle_2$; or

$$U_L\left[|1\rangle_2\left(\alpha_{01}|0\rangle_3|\varepsilon_{01}\rangle+\alpha_{11}|1\rangle_3|\varepsilon_{11}\rangle\right)\right]=|1\rangle_2|1\rangle_3|\varepsilon_1\rangle.\tag{30}$$

when Bob's measurement result is $|1\rangle_2$.

(ii) Case 2: Bob chooses the CTRL operation. As a result, the global state of the composite system is $\frac{1}{\sqrt{2}}\left[|0\rangle_2\left(\alpha_{00}|0\rangle_3|\varepsilon_{00}\rangle+\alpha_{10}|1\rangle_3|\varepsilon_{10}\rangle\right)+|1\rangle_2\left(\alpha_{01}|0\rangle_3|\varepsilon_{01}\rangle+\alpha_{11}|1\rangle_3|\varepsilon_{11}\rangle\right)\right]$.

Eve performs $U_L$ on the particle back to Alice. According to Eq.(29) and Eq.(30), it can be obtained that

$$\begin{aligned}U_L\left[U_H\left(|\phi^+\rangle_{23}|\varepsilon\rangle_E\right)\right]&=\frac{1}{\sqrt{2}}U_L\left[|0\rangle_2\left(\alpha_{00}|0\rangle_3|\varepsilon_{00}\rangle+\alpha_{10}|1\rangle_3|\varepsilon_{10}\rangle\right)+|1\rangle_2\left(\alpha_{01}|0\rangle_3|\varepsilon_{01}\rangle+\alpha_{11}|1\rangle_3|\varepsilon_{11}\rangle\right)\right]\\&=\frac{1}{\sqrt{2}}\left(|0\rangle_2|0\rangle_3|\varepsilon_0\rangle+|1\rangle_2|1\rangle_3|\varepsilon_1\rangle\right)\\&=\frac{1}{2}\left[\left(|\phi^+\rangle_{23}+|\phi^-\rangle_{23}\right)|\varepsilon_0\rangle+\left(|\phi^+\rangle_{23}-|\phi^-\rangle_{23}\right)|\varepsilon_1\rangle\right]\\&=\frac{1}{2}\left[|\phi^+\rangle_{23}\left(|\varepsilon_0\rangle+|\varepsilon_1\rangle\right)+|\phi^-\rangle_{23}\left(|\varepsilon_0\rangle-|\varepsilon_1\rangle\right)\right].\end{aligned}\tag{31}$$

In order that Eve induces no error on the CTRL particle in Step 7, Alice's Bell basis measurement result on the particle in $S_2^{'U}$ reflected by Bob and the corresponding particle in $S_3^{'}$ at her site should be in the state of $|\phi^+\rangle_{23}$. Thus, we should have

$$|\varepsilon_0\rangle=|\varepsilon_1\rangle=|\varepsilon\rangle.\tag{32}$$

(iii) Inserting Eq.(32) into Eqs.(29-31) produces

$$U_L\left[|0\rangle_2\left(\alpha_{00}|0\rangle_3|\varepsilon_{00}\rangle+\alpha_{10}|1\rangle_3|\varepsilon_{10}\rangle\right)\right]=|0\rangle_2|0\rangle_3|\varepsilon\rangle,\tag{33}$$

$$U_L\left[|1\rangle_2\left(\alpha_{01}|0\rangle_3|\varepsilon_{01}\rangle+\alpha_{11}|1\rangle_3|\varepsilon_{11}\rangle\right)\right]=|1\rangle_2|1\rangle_3|\varepsilon\rangle,\tag{34}$$

$$U_L\left[U_H\left(|\phi^+\rangle_{23}|\varepsilon\rangle_E\right)\right]=|\phi^+\rangle_{23}|\varepsilon\rangle,\tag{35}$$

respectively.

It can be concluded from Eqs.(33-35) that in order to introduce no error in Step 7, the final state of Eve's probe should be independent from both Bob's operation and Alice and Bob's measurement results.

(3) The double CNOT attacks

In Step 5, Alice sends the particles of $S_2^{'U}$ to Bob one by one. After intercepting the particle from Alice to Bob, Eve performs the first CNOT attack on her own auxiliary qubit $|0\rangle_E$ and it. Here, her own auxiliary qubit $|0\rangle_E$ acts as the target qubit, which the intercepted particle plays the role of the control qubit. After the first CNOT operation from Eve, the global state of the whole system can be depicted as

$$CNOT_{2E}\left(\left|\phi^{+}\right\rangle_{23}\otimes\left|0\right\rangle_{E}\right)=\frac{1}{\sqrt{2}}\left(\left|000\right\rangle+\left|111\right\rangle\right)_{23E}. \tag{36}$$

Then, Eve sends the particle of $S_2'^U$ to Bob. In Step 6, the received particle is randomly performed with the CTRL operation or the SIFT operation from Bob. Eve intercepts the particle back to Alice. Since Eve has no knowledge about Bob's operation, for escaping from being discovered, Eve has to implement the second CNOT attack on the particle back to Alice and her auxiliary qubit, where the former and the latter are the control qubit and the target qubit, respectively. Two cases need to be considered.

Case 1: Bob implements the CTRL operation. After the second CNOT attack from Eve, the global state of the whole system is turned into

$$CNOT_{2E}^{\otimes 2}\left(\left|\phi^{+}\right\rangle_{23}\otimes\left|0\right\rangle_{E}\right)=\frac{1}{\sqrt{2}}\left(\left|000\right\rangle+\left|110\right\rangle\right)_{23E}=\left|\phi^{+}\right\rangle_{23}\left|0\right\rangle_{E}. \tag{37}$$

Case 2: Bob implements the SIFT operation. As a result, the global state of the whole system in Eq.(36) will be collapsed into $\left|000\right\rangle_{23E}$ or $\left|111\right\rangle_{23E}$. After the second CNOT attack from Eve, the global state of the whole system becomes

$$CNOT_{2E}\left(\left|000\right\rangle\right)_{23E}=\left|00\right\rangle_{23}\left|0\right\rangle_{E}, \quad \text{if Bob's measurement result is}\left|0\right\rangle_{2}; \tag{38}$$

$$CNOT_{2E}\left(\left|111\right\rangle\right)_{23E}=\left|11\right\rangle_{23}\left|0\right\rangle_{E}, \quad \text{if Bob's measurement result is}\left|1\right\rangle_{2}. \tag{39}$$

According to Eqs.(37-39), although Eve can escape the second eavesdropping detection in Step 7 after her double CNOT attacks, he still has no knowledge about both Bob's operation and Alice and Bob's measurement results, since Eve's auxiliary particle after her double CNOT attacks is always in the state of $\left|0\right\rangle_{E}$. It can be concluded that this protocol can overcome this kind of double CNOT attacks.

(4) The measure-resend attack

In Step 5, after Alice performs the corresponding unitary operation on the particle of $S_2'$, the state of the particles of the same position in $S_2'^U$ and $S_3'$ is always in the state of $\left|\phi^{+}\right\rangle$. After intercepting the particles in $S_2'^U$ from Alice to Bob, in order to try to get $M_{A2}$, Eve measures them with the $Z$ basis. As a result, each of the particles in $S_2'^U$ is randomly collapsed into $\left|0\right\rangle_{2}$ or $\left|1\right\rangle_{2}$ with the same probability. Without loss of generality, suppose that the $i$ th particle in $S_2'^U$ is collapsed into $\left|0\right\rangle_{2}$. When Bob chooses the CTRL operation, after Bob tells Alice her operation, Alice measures the particle of $S_2'^U$ from Bob and the corresponding particle in $S_3'$ in her hand with the Bell basis; as a result, Eve is detected by Alice and Bob with the probability of $\frac{1}{2}$ in this case, as Alice's measurement result is randomly in $\left|\phi^{+}\right\rangle_{23}$ or $\left|\phi^{-}\right\rangle_{23}$. When Bob chooses the SIFT operation, after Bob tells Alice her operation, Alice will measure the particle of $S_2'^U$ from Bob and the corresponding particle in $S_3'$ in her hand with the $Z$ basis if this position is chosen for the second eavesdropping detection; as a result, Eve is detected by Alice and Bob with the probability of 0 in this case, as Alice's measurement results on the particle of $S_2'^U$ from Bob and the corresponding particle in $S_3'$ in her hand are $\left|0\right\rangle_{2}$ and $\left|0\right\rangle_{3}$, respectively.

To sum up, when Eve initiates this measure-resend attack, the probability of Eve being detected in the second eavesdropping detection is $\frac{1}{2} \times \frac{1}{2} + \frac{1}{2} \times 0 = \frac{1}{4}$.

(5) The intercept-resend attack

In order to try to obtain $M_{A2}$, Eve intercepts the particles in $S_2'^U$ sent from Alice to Bob, and sends the fake ones she prepared in the $Z$ basis in advance to Bob. Without loss of generality, suppose that the $i$ th fake particle Eve prepares is $|0\rangle_E$. After intercepting the $i$ th particle in $S_2'^U$ sent from Alice, Eve sends the $i$ th fake particle $|0\rangle_E$ to Bob. When Bob chooses the CTRL operation, after Bob tells Alice her operation, Alice measures the $i$ th fake particle $|0\rangle_E$ from Bob and the $i$ th corresponding particle in $S_3'$ in her hand with the Bell basis; as a result, Eve is detected by Alice and Bob with the probability of $\frac{3}{4}$ in this case, as Alice's measurement result is randomly in $|\phi^+\rangle_{E3}$, $|\phi^-\rangle_{E3}$, $|\psi^+\rangle_{E3}$ or $|\psi^-\rangle_{E3}$. When Bob chooses the SIFT operation, after Bob tells Alice her operation, Alice will measure the $i$ th fake particle $|0\rangle_E$ from Bob and the $i$ th corresponding particle in $S_3'$ in her hand with the $Z$ basis if this position is chosen for the second eavesdropping detection; as a result, Eve is detected by Alice and Bob with the probability of $\frac{1}{2}$ in this case, as Alice's measurement result on the $i$ th corresponding particle in $S_3'$ in her hand is randomly $|0\rangle_3$ or $|1\rangle_3$.

All in all, when Eve executes the intercept-resend attack, the probability of Eve being detected in the second eavesdropping detection is $\frac{1}{2} \times \frac{3}{4} + \frac{1}{2} \times \frac{v}{n+v} \times \frac{1}{2} = \frac{3}{8} + \frac{v}{4(n+v)}$.

## 4 Discussions and conclusions

The qubit efficiency, defined by Cabello [36] as

$$\eta = \frac{\lambda_b}{\gamma_q + \gamma_c}, \qquad (40)$$

is often used to evaluate the performance of efficiency for a quantum communication protocol, where $\lambda_b$, $\gamma_q$ and $\gamma_c$ are the number of the private shared key bits, the number of consumed qubits, and the number of classical bits used for the classical communication, respectively. In the following, when calculating the qubit efficiency, we disregard the classical bits consumed for security checks.

In the proposed SQKD protocol, Alice and Bob successfully establish an INFO bit sequence of length $3n+2v$, so $\lambda_b = 3n+2v$. Alice needs to prepare $4(n+\delta+v)$ GHZ-like states all in the state of $|G_{001}\rangle$ and sends the particles in $S_1$ to Bob; when Bob chooses the SIFT operations for half of the received particles in $S_1$, he needs to produce $2(n+\delta+v)$ new particles in the states as found within the $Z$ basis; when Bob chooses the SIFT operations for half of the received particles in $S_2'^U$, he needs to produce $n+v$ new particles in the states as found within the $Z$ basis; hence, we have $\gamma_q = 4(n+\delta+v) \times 3 + 2(n+\delta+v) + (n+v) = 15n + 14\delta + 15v$. There are no classical bits consumed for the classical communication, so $\gamma_c = 0$. Consequently, we have $\eta = \frac{3n+2v}{15n+14\delta+15v}$.

In addition, we make a detailed comparison among the existing SQKD protocols based on three-particle entangled states, and summarize the comparison results in Table 1.

Table 1  Comparison results of the existing SQKD protocols based on three-particle entangled states

| | The first protocol of Ref.[27] | The second protocol of Ref.[27] | The first protocol of Ref.[28] | The second protocol of Ref.[28] | This protocol |
|---|---|---|---|---|---|
| Function | Establishing a private shared key between one quantum party and one classical party under the help of a quantum third party | Establishing a private shared key between two classical parties under the help of a quantum third party | Establishing a private shared key between one quantum party and one classical party without the existence of a third party | Establishing a private shared key between two classical parties under the help of a quantum third party | Establishing a private shared key between one quantum party and one classical party without the existence of a third party |
| Whether exists a third party | Yes | Yes | No | Yes | No |
| Initial quantum resource | GHZ states and sing-qubit states | GHZ states and sing-qubit states | GHZ-like states | GHZ-like states | GHZ-like states |
| Quantum measurements from the quantum party | The $Z$ basis measurements and the $X$ basis measurements | The $Z$ basis measurements and the $X$ basis measurements | The $Z$ basis measurements and the Bell basis measurements | The $Z$ basis measurements and the GHZ-like basis measurements | The $Z$ basis measurements, the Bell basis measurements and the GHZ-like basis measurements |
| Quantum measurements from the classical party | The $Z$ basis measurements | The $Z$ basis measurements | The $Z$ basis measurements | The $Z$ basis measurements | The $Z$ basis measurements |
| Usage of pauli operations | No | No | No | No | Yes |
| Usage of Hadamard gate | No | Yes | No | No | No |
| Usage of quantum entanglement swapping | No | No | No | No | No |

To sum up, in this paper, in order to realize the aim of establishing a private shared key between one quantum party and one classical party without the existence of a third party, we design a novel two-party SQKD protocol by only using one kind of GHZ-like state as the initial quantum resource. The proposed protocol employs neither the Hadamard gate nor quantum entanglement swapping. We validate that it can overcome the Trojan horse attacks, the entangle-measure attack, the double CNOT attacks, the measure-resend attack and the intercept-resend attack.

## Acknowledgments

Funding by the National Natural Science Foundation of China (Grant No.62071430 and No.61871347) and the Fundamental Research Funds for the Provincial Universities of Zhejiang (Grant No.JRK21002) is gratefully acknowledged.